\documentclass[twocolumn]{emulateapj}
\usepackage{amsmath}
\usepackage{natbib}

\newcommand{\Msolar}{M$_{\odot}$}
\newcommand{\Lsolar}{L$_{\odot}$}

\begin{document}

\title{Different Dynamical Ages for the Two Young and Coeval LMC Star Clusters NGC~1805 and NGC~1818 Imprinted on Their Binary Populations}
\shorttitle{Different Dynamical Ages for NGC~1805 and NGC~1818}

\author{Aaron M.~Geller$^{a,b}$}
\affil{Center for Interdisciplinary Exploration and Research in Astrophysics (CIERA) and Department of Physics and Astronomy, Northwestern University, 2145 Sheridan Road, Evanston, IL 60208, USA}
\affil{Department of Astronomy and Astrophysics, University of Chicago, 5640 S. Ellis Avenue, Chicago, IL 60637, USA}
\email{$^a$ a-geller@northwestern.edu}
\thanks{$^b$ NSF Astronomy and Astrophysics Postdoctoral Fellow}

\author{Richard de Grijs}
\affil{Kavli Institute for Astronomy and Astrophysics, Peking University, Yi He Yuan Lu 5, Hai Dian District, Beijing 100871, China}
\affil{Department of Astronomy, Peking University, Yi He Yuan Lu 5, Hai Dian District, Beijing 100871, China}

\author{Chengyuan Li}
\affil{Department of Astronomy, Peking University, Yi He Yuan Lu 5, Hai Dian District, Beijing 100871, China}
\affil{Kavli Institute for Astronomy and Astrophysics, Peking University, Yi He Yuan Lu 5, Hai Dian District, Beijing 100871, China}

\author{Jarrod R.~Hurley}
\affil{Centre for Astrophysics and Supercomputing, Swinburne University of Technology, VIC 3122, Australia}

\shortauthors{Geller et al.}

\begin{abstract}

The two Large Magellanic Cloud star clusters NGC~1805 and NGC~1818 are approximately the same chronological age ($\sim$30~Myr), 
but show different radial trends in binary frequency.  The F-type stars (1.3 - 2.2~\Msolar) in NGC~1818 have a binary frequency that decreases 
towards the core, while the binary frequency for stars of similar mass in NGC~1805 is flat with radius, or perhaps bimodal (with a peak in the core).  
We show here, through detailed $N$-body modeling, that both clusters could have formed with the same primordial binary frequency and with binary orbital 
elements and masses drawn from the same distributions (defined from observations of open clusters and the field of our Galaxy).  The observed radial 
trends in binary frequency for both clusters are best matched with models that have initial substructure.  Furthermore, both clusters may be evolving 
along a very similar dynamical sequence, with the key difference that NGC~1805 is dynamically older than NGC~1818.  The F-type binaries in NGC~1818 still 
show evidence of an initial period of rapid dynamical disruptions (which occur preferentially in the core), while NGC~1805 has already begun to recover 
a higher core binary frequency, owing to mass segregation (which will eventually produce a distribution in binary frequency that rises only towards the core, as is 
observed in old Milky Way star clusters).  This recovery rate increases for higher-mass binaries, and therefore even at one age in one cluster, we predict a similar 
dynamical sequence in the radial distribution of the binary frequency as a function of binary primary mass.

\end{abstract}


\section{Introduction}\label{intro}

Binary, and higher-order multiple, stars are ubiquitous, and comprise a relatively large fraction of the stars in star forming regions \citep{ghe93,koh00,kra11,bat12,san13}, open clusters 
\citep{mer92,pat98,pat02,gel08,gel10,hol09} and the Galactic field \citep{rag10}.  
For solar-type stars, the binary frequency\footnote{We follow the usual convention and define the binary frequency as $f_{\rm b} = N_{\rm b} / (N_{\rm s} + N_{\rm b} + ...)$, where $N_{\rm b}$ is the number of binaries, 
$N_{\rm s}$ is the number of single stars, and ``...'' signifies higher-order multiples.} in the Galactic field is approximately 50\%, while
for the most massive stars, the binary frequency increases to about 70\% \citep[and perhaps even higher, e.g.][]{rag10,san12,cab14}.  

Binaries in the Galactic field live essentially in isolation, and only the very widest field binaries (e.g., those with separations over $\sim$10$^3$ AU) are
in danger of encountering passing stars, or having their orbits changed dramatically by the Galactic tidal field \citep{kai14}.
However, most stars (with masses $\geq$0.5~\Msolar) form in denser environments \citep{lad03,eva09,bre10}.  
Though many young, embedded, star clusters quickly dissolve to populate the Galactic field \citep{ada01}, close stellar 
encounters within these birth environments are capable of significantly modifying the orbits of, and even disrupting, binary systems that would otherwise be 
stable in the Galactic field. Therefore, our interpretation of the observed binary populations in star clusters and the field, as well as our understanding of star 
formation in general, relies on how a population of stars and binaries evolves through this more dynamically active early stage in a star cluster.

Within star clusters, binaries are typically discussed as either being ``hard'' or ``soft'' \citep{heg75}.  Hard binaries have high binding energies relative to 
the kinetic energies of stars moving throughout the cluster, and therefore encounters tend to make hard binaries more tightly bound, or harder.
The converse is true for soft binaries, and indeed an encounter involving a soft binary can completely unbind the system.  

Observations of binary populations in young star clusters are very valuable for our understanding of the early phases of star cluster dynamics.
Two of the youngest well-sampled star clusters that have such observations are the Large Magellanic Cloud (LMC) star clusters NGC~1805 and NGC~1818, both with an age of approximately 30 Myr \citep{deg02,li13}\footnote{
\citet{deg02} estimate an age for NGC~1805 of $\log (t$ [yr]$) = 7.00 \substack{+0.30 \\ -0.10}$ and for NGC~1818 of $7.40 \substack{+0.30 \\ -0.10}$ (from a thorough review of the prior literature), while \citet{li13} 
find an age for NGC~1805 of $\log (t$ [yr]$) = 7.65 \pm 0.10$ and for NGC~1818 of $7.25 \pm 0.10$.  This slight disagreement in the cluster ages likely is a result of improved isochrones (and 
better field star decontamination) employed by \citet{li13}.  For our modeling we use a single age of 30 Myr for both clusters, which is well within a 3$\sigma$ confidence interval from each of these age estimates.}.
For both star clusters, \citet{deg13} and \citet{li13} measure the binary frequency as a function of radius from the cluster center.  Interestingly, they find that the two 
clusters show different trends, 
where the radial distribution of the binary frequency in NGC~1818 decreases towards the cluster core, and the distribution in NGC~1805 is flat or perhaps bimodal 
(with a peak in the core).

Their result for NGC~1818 is particularly interesting because, typically, the binary frequencies of star clusters are observed to have 
the opposite radial trend, increasing towards the core of the cluster \citep[e.g.][]{gel12,mil12}.  
This phenomenon is understood to be a result of mass segregation, 
where, in a sample containing binaries with primary stars of similar masses to the single stars, the more massive binaries sink towards the cluster center, 
due to two-body relaxation and dynamical friction.

In \citet{gel13b}, we studied NGC~1818 in detail through $N$-body modeling.  We found that, for a cluster born with soft binaries and no radial dependence of the
binary frequency, the observed trend of the binary frequency decreasing towards the cluster core can be naturally explained through the early disruption of wide binaries
by close encounters with other stars
(on approximately a crossing time).  This process occurs preferentially towards the denser core of the cluster, which explains the lower binary frequency in the core relative to the halo.  
Over a few relaxation times, the binary frequency evolves under the influence 
of mass-segregation effects, to produce a bimodal radial distribution, and eventually a distribution that rises only towards the core (as is often observed in older star clusters).

The observed bimodal-like radial trend in binary frequency for NGC~1805 \citep{li13} is similar to the second phase in this evolutionary sequence, where mass-segregation effects begin to become important 
for determining the radial distribution of the binary frequency.
However, NGC~1805 and NGC~1818 are approximately the same chronological age,
and furthermore, our models of NGC~1818 suggest that even taking the extremes of the age range from \citet[][and references therein]{deg02} and \citet{li13}, cannot alone explain the 
differences in their binary frequency radial distributions (with all else being equal between the clusters).

The two key differences between these clusters, from a dynamical evolution perspective, is that NGC~1805 is less massive and more centrally concentrated than NGC~1818.
\citet{mac03} estimate that NGC~1805 and NGC~1818 have masses of $\log (M/$\Msolar$) = 3.52 \pm 0.13$ and $4.13 \substack{+0.15 \\ -0.14}$, respectively, and 
core radii of $1.33 \pm 0.06$~pc and $2.45 \pm 0.09$~pc, respectively.
Consequently, \citet{deg02} estimate that the current half-mass relaxation time in NGC~1805 is 4 to 5 times shorter than in NGC 1818 \citep{deg02}.

Here we use $N$-body modeling to investigate whether the observations of both NGC~1805 and NGC~1818 
can be reproduced with similar initial binary populations, all undergoing the evolutionary sequence discussed in \citet{gel13b}.
If so, this may suggest that binaries can form with similar properties across different environments in the LMC, as is also discussed in \citet{gel13} in the context of 
open clusters and the field in our Galaxy.

We briefly discuss the setup of our simulations in Section~\ref{method}.
In Section~\ref{simtoobs} we compare the models directly to the observations of NGC~1805.  We then expand upon the analysis 
of \citet{gel13b} and study the evolution of the binary frequency as a function of binary primary-star mass in Section~\ref{phenom}.
We offer our explanation of the different observed trends in the radial dependence of the binary frequency in NGC~1805 and NGC~1818,
through comparisons to our $N$-body models, in Section~\ref{compto1818}.  Finally, in Sections~\ref{discuss}~and~\ref{conc} we provide 
a brief discussion and conclusions.

\section{$N$-body Simulations} \label{method} 

We perform a modest grid of direct $N$-body simulations using the \texttt{NBODY6} code \citep{aar03} targeted at reproducing the observed 
surface density profile of NGC~1805 from \citet{mac03} at an age of 30 Myr.   
Our modeling procedure here is nearly identical to that of \citet{gel13,gel13b}, and therefore we point the reader to those papers for additional details not discussed here.  

We stress that the initial conditions for the binary populations in these NGC~1805 models are nearly identical to those of our NGC~1818 
models from \citet{gel13b}.  Moreover, we input a 100\% initial binary frequency (independent of stellar mass and radial position), with binary orbital 
parameters distributed according to observations of solar-type binaries in young open clusters and the Galactic field \citep[e.g.][]{gel10,rag10}.
Most importantly for our later discussion, the initial orbital period distribution is log-normal (as observed for solar-type binaries, 
with a mean of $\log (P$~[days])~=~5.03 and $\sigma$~=~2.28), and extends from 
periods of about 0.1 days to $10^{10}$ days.  As explained in 
\citet[][see particularly Figures 3 and 4]{gel13b},
and below, many of the very wide (soft) binaries are disrupted quickly by encounters 
with other stars.  The initial orbital eccentricities are drawn from a Gaussian distribution with a mean of $e = 0.38$ with $\sigma = 0.23$,
as is observed for solar-type binaries in the young ($\sim$150 Myr) open cluster M35 and consistent with similar binaries in the Milky Way field \citep{gel10,gel13,rag10}.
We draw binary mass ratios from a uniform distribution, but with limits such that the mass-ratio is always less than unity and the mass of the secondary star is always 
greater than 0.1 \Msolar, which produces a mass-ratio distribution that is approximately of the form $\rm d N/\rm d q \propto q^{-0.4}$, within the observed mass range 
of \citet{li13}; see \citealt{gel13b} for further details.
The only differences that are imposed on the initial binaries in these NGC~1805 models from our NGC~1818 models arise from the differences in observed 
cluster masses and structures, as we explain below.

We choose to investigate two different initial virial ratios of $Q = 0.5$ (equilibrium) and $Q = 0.3$ (collapsing), and two different degrees of 
substructure, using fractal distributions of degrees\footnote{
We follow \citet{goo04} and \citet{kup11} to define the degree of substructure with the parameter, $D$, such that the probability that a given particle will have a nearby ``child'' is $2^{(D-3)}$, where
$D = 3$ produces no substructure and $D < 3$ produces fractal density distributions.  We refer the reader to \citet{goo04} and \citet{kup11} for specific details on the algorithm used by \texttt{McCluster} 
for defining the initial fractal density distributions}
$D = 3$ (smooth) and $D = 2$ (clumpy).  We follow the same procedure as in \citet{gel13b} to generate
these initial conditions, namely smooth distributions are generated within the \texttt{NBODY6} code and substructured initial conditions are generated 
using \texttt{McLuster} \citep{kup11}.  
In short, to set up the clumpy models, \texttt{McLuster} first defines a fractal distribution of stars within a unit sphere, and 
then ``folds'' this distribution with a given density profile.
For simplicity, we choose to begin all models with an underlying Plummer density profile \citep{plu11}; substructure is imposed on top of the Plummer model for 
those simulations with $D=2$.  Thus, all models begin centrally concentrated, as defined by the Plummer scale radius (or equivalently, the virial radius, as discussed below).
At an age of 30 Myr in the simulations, our initially substructured models relax to have smooth density distributions
(e.g., see Figure~1 of \citealt{gel13b}).  

Likely there are other potential combinations of initial $Q$ and $D$ that could reproduce the observations of NGC~1805 at an age of 30 Myr.  We choose here to 
investigate [$Q$,$D$] = [0.5,~3] as a baseline model to compare against collapsing ($Q < 0.5$) and substructured ($D < 3$) models.  In general, other studies (including 
our modeling of NGC 1818) have simulated star clusters with $Q$ ranging from about 0.1 to 0.75, where $Q > 0.5$ are initially expanding clusters. 
Observations suggest that many clusters form with subvirial ($Q < 0.5$) velocities \citep{per06,and07,tob09,pro09}, and therefore we choose to focus on subvirial models and  investigate a moderate value of $Q = 0.3$. 
Likewise, other studies have investigated a range in $D$, from about 1 to 3.  Again, here we choose a moderate value of $D = 2$ for our substructured models.
We do not attempt to identify the most probable specific initial values of [$Q$,$D$] through our simulations (and indeed this is likely not possible to any relevant precision, given the observations); 
instead we investigate for differences between our baseline (smooth, equilibrium) model and modestly subvirial and/or substructured models.

Compared to NGC~1818, NGC~1805 is relatively compact.  The  \citet[][EFF]{els87} profile fit to the observations of NGC~1805 by \citet{mac03} gives 
$a$~=~6.84~$\pm$~0.42~arcsec 
(1.66~$\pm$~0.10~pc, using the canonical LMC distance modulus of 18.5, which equates to a scale of 4.116~arcsec~pc$^{-1}$)
and $\gamma$~=~2.81~$\pm$~0.10, which can be converted into a King core radius of 1.33~pc.  
For reference a similar calculation for NGC~1818 yields a core radius of 2.45~pc.
\citet{mac03} found a central surface brightness in NGC~1805 of $\log \mu_0$~=~3.49~$\pm$~0.02~\Lsolar~pc$^{-2}$
(which implies a central mass surface density of 155~$\pm$~7~\Msolar~pc$^{-2}$, given the mass-to-light ratio of 0.05 adopted by \citealt{mac03}).
The implied total mass of NGC~1805 is then $\log (M/$\Msolar$) = 3.52 \pm 0.13$.
For reference, \citet{deg02} found a mass of NGC~1805 of $2.8\substack{+3.0 \\ -0.8} \times 10^3$~\Msolar, while \citet{joh01} found a somewhat
higher mass of 6000~\Msolar.

We aim to create a model of NGC~1805 that reasonably matches these parameters at an age 30 Myr, by adjusting the initial virial radius (or equivalently, the Plummer scale radius), 
and initial total mass.  From comparisons to our previous simulations of other star clusters \citep{hur05,gel13,gel13b},  we predict that only a small amount of 
mass loss will occur over the 30 Myr timescale of interest.
To begin, we ran a series of trial simulations of NGC~1805 over a grid in initial number of stars and virial radius, and determined that starting with 7600 stars, 
drawn from a \citet{kro01} initial-mass function (with masses between 0.1~\Msolar\ and 50~\Msolar) will evolve to be within the observed mass range of NGC~1805 at 30 Myr.
This initial number of stars and mass function results in an initial total cluster mass of $\sim$4700~\Msolar~(already within the range of masses estimated from the observations of NGC~1805, though 
towards the high end).
The true initial number of stars, and corresponding initial total cluster mass, of NGC~1805 is most likely not exactly what we have chosen,
In general, beginning with more stars (with all else being equal) 
would result in a longer initial half-mass relaxation time, and therefore less rapid dynamical evolution.  The opposite is true for a cluster with a lower number of initial stars.
As we discuss below, our simulations
do indeed reside within the observed mass range of NGC~1805 at an age of 30 Myr, which is sufficient for our purposes.  

To determine the initial virial radius for a given model, we compared results from our initial set of trial simulations to the observed surface density profile 
at an age of 30 Myr.  We find initial virial radii between 4 and 6 pc (or equivalently Plummer scale radii between 2.36 and 3.53 pc)
fit well to the observations at 30 Myr.
Specifically we will present models with initial virial ratios, fractal dimension, and virial radii, 
[$Q$,~$D$,~$R_V$~(pc)] = [0.5,~3,~4], [0.5,~2,~4], [0.3,~3,~4], [0.3,~2,~6].
We do not model the embedded phase of the cluster here, and instead begin our simulations at $t=0$ after gas expulsion
and with all stars on the zero-age main sequence.
We compare the simulations at 30 Myr to the observations in Section~\ref{simtoobs}.

\begin{figure}[!t]
\plotone{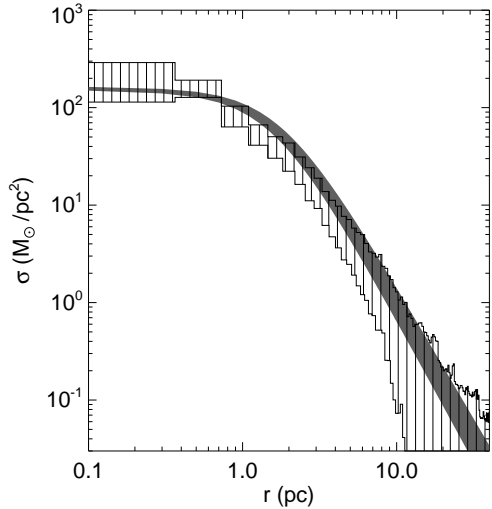}
\caption{
Projected radial mass surface density profile for one NGC~1805 simulation compared to the EFF profile fit to the observations of the cluster by \citet{mac03}.
We show the same [$Q$,$D$] = [0.5,~2] simulation here as in Figure~\ref{bfreqvr}, which matches the observed radial distribution of the binary frequency particularly closely.  
(All of our simulations reproduce the observed surface density profile.)  The binned and hatched area shows the region from the simulations within which fall 95\% of the 1000
random sight lines. The solid gray band shows the region encompassed by the \citet{mac03} EFF model, with parameters 
$\log \mu_0$~=~3.49~$\pm$~0.02~\Lsolar~pc$^{-2}$ (155~$\pm$~7~\Msolar~pc$^{-2}$),
$a$~=~1.66~$\pm$~0.10~pc,
$\gamma$~=~2.81~$\pm$~0.10.
\label{sdensity}
}
\end{figure}

To simplify our comparison between models of NGC~1805 and NGC~1818, we place these simulations of NGC~1805 in the same orbit around the LMC as we used in \citet{gel13b} for 
the NGC~1818 simulations.  Specifically we start the cluster at 3.3 kpc from the center of a point mass of $10^{10}$~\Msolar\ on a linearized circular orbit.
NGC~1805 is observed to be at about 3.86 - 4.00 degrees \citep[$\sim$3.4 - 3.5 kpc][]{mac03} from the center of the LMC, but little more is known about the cluster's 
orbit.  This, and the very minor effect on the cluster from the LMC tidal field over 30 Myr, suggests that this small difference between the observations and our choice of
orbit is unimportant for the present analysis and understanding.
We also note that, in addition to both NGC~1805 and NGC~1818 being at a similar distance from the center of the LMC, they are also on the same side of the parent galaxy, 
and therefore are also affected in a similar manner by the tidal field of the Small Magellanic Cloud (though this minor effect is not modeled here).

We produce 20 simulations at each combination of virial ratio ($Q$) and fractal dimension ($D$), drawing from the same initial parameter distributions but varying 
the initial random seed, in order to address the stochasticity in $N$-body simulations.\footnote{It is a fairly standard practice to run at least 10 simulations, with initial 
conditions drawn from the same parent distributions, to assess the expected variation 
among $N$-body models of a given star cluster \citep[e.g.,][]{par12,kou14}.}
In total we present results from 80 simulations of NGC~1805.

\begin{figure}[!t]
\plotone{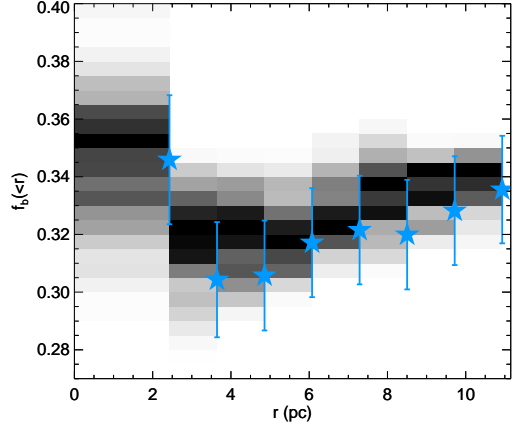}
\caption{
Binary frequency as a function of radius from the cluster center, comparing one NGC~1805 simulation to the observations of \citet{li13}.
For both the observations and the simulation, all single stars and binaries with primary-star masses between 1.3~\Msolar\ and 2.2~\Msolar\ are included.
For binaries, we only count those with a mass ratio ($q=m_2/m_1$, where $m_1 > m_2$) $\geq$0.55, and each binary with $q < 0.55$ is counted as one single star 
(due to observational limitations).
The observations are shown in colored stars with error bars, and indicate the total binary frequency inside the radius where the point is located.
In the gray-scale distribution, we show the results from the same [$Q$,$D$] = [0.5,~2] simulation as in Figure~\ref{sdensity}.
We use the same cumulative accounting of the binary frequency at the same radial locations as the observations,
but for the simulation, we plot the distribution of the binary frequency at each radius resulting from the 1000 different projected sight lines (in bins of $f_b$ = 0.005, 
with darker values indicating higher likelihood).
This simulation very closely reproduces the observed bimodal distribution in binary frequency as a function of radius.
\label{bfreqvr}
}
\end{figure}

\begin{deluxetable*}{ccccccccc}[!t]
\tablecaption{Summary Table of $N$-body Simulations\label{tab1}}
\tablehead{\colhead{$Q$} & \colhead{$D$} & \colhead{$R_{\rm V}$} & \colhead{$N_{\rm sims}$} & \colhead{$M_c$} & \colhead{$\sigma_{\rm 0}$(obs)} & \colhead{$f_{\rm b}$(tot)} & \colhead{$f_{\rm b}$(obs)} & \colhead{$P_{\chi^2}$(obs)} \\
\colhead{} & \colhead{} & \colhead{(pc)} & \colhead{} & \colhead{ (\Msolar)} & \colhead{(\Msolar\ pc$^{-2}$)} & \colhead{} & \colhead{} & \colhead{(\%)}}
0.5 & 3 & 4 & 20 & 4092~$\pm$~12 & 105~$\substack{+70 \\ -61}$ & 0.7637~$\pm$~0.0023 & 0.29~$\substack{+0.04 \\ -0.04}$ &  27.840 \\
0.5 & 2 & 4 & 20 & 4017~$\pm$~16 & 148~$\substack{+109 \\ -88}$ & 0.7450~$\pm$~0.0022 & 0.30~$\substack{+0.04 \\ -0.07}$ &  59.315 \\
0.3 & 3 & 4 & 20 & 4098~$\pm$~12 & 121~$\substack{+71 \\ -60}$ & 0.7578~$\pm$~0.0024 & 0.28~$\substack{+0.03 \\ -0.04}$ &  34.030 \\
0.3 & 2 & 6 & 20 & 4099~$\pm$~20 & 108~$\substack{+143 \\ -70}$ & 0.8004~$\pm$~0.0024 & 0.31~$\substack{+0.06 \\ -0.08}$ &  49.610 \\

\enddata
\end{deluxetable*}

\section{Comparison Between Observations and Simulations of NGC~1805} \label{simtoobs}

All of the simulations listed in Section~\ref{method} 
evolve to be consistent with the 
observed total mass, total binary frequency and surface density profile of NGC~1805 at an age of 30 Myr.  
As discussed above, NGC~1805 is observed to have a mass between 2000~\Msolar\ and 6000~\Msolar\ (and \citealt{mac03} derived a mass
between 2450~\Msolar\ and 4470~\Msolar).  At 30 Myr our models range in mass from about 3838~\Msolar\ to 4312~\Msolar.
\citet{li13} found a binary fraction for NGC~1805 between about 32\% and 40\% for binaries with primary masses between 1.3~\Msolar\ and 2.2~\Msolar, mass ratios $\geq$0.55 and within a
radius of 45 arcsec (or about 11 pc at the distance of the LMC) from the cluster center.
Within the same observational limits, our simulations have binary frequencies that range from about 21\% to 36\% (and total binary frequencies between 73\% and 82\%), at 30 Myr. 

In Figure~\ref{sdensity} we show one example comparison to the observed surface density profile of NGC~1805 for a particular simulation that also closely matches the observed radial 
distribution of the binary frequency (see Figure~\ref{bfreqvr}).  
Here, as we will also do throughout the paper when analyzing simulations in projection, we take 1000 different randomly chosen lines-of-sight through our model
and combine the results.  
Strictly, the orbit of the simulated cluster within the LMC potential defines a true line-of-sight projection for the cluster relative
to an observer on Earth.  However, given the uncertainties in the true cluster orbital parameters, our simplistic modeling of the
LMC potential and random asymmetries within the initial conditions of substructured simulations, we prefer to select many random lines-of-sight 
and analyze the resulting distributions as an ensemble.
In general we find that the results (e.g., surface density profile, radial distribution of the binary frequency, 
etc.) are not particularly sensitive to the specific choice of projection.  

We also provide summary information for our simulations in Table~\ref{tab1}, including the initial virial ratio ($Q$), fractal dimension ($D$), and virial radius ($R_{\rm V}$), the
number of simulations with these initial parameters ($N_{\rm sims}$), and the 30 Myr total cluster mass ($M_c$), central surface density ($\sigma_{\rm 0}$(obs), only including stars with $V$ magnitudes $< 25$, approximately the faint limit of observations from \citealt{mac03}),
total binary frequency ($f_{\rm b}$(tot)), observable binary frequency ($f_{\rm b}$(obs), for stars in the simulations within the observed range of \citealt{li13}, stated above) and the percentage of simulations that 
match the observed binary frequency radial distribution of \cite{li13} to within $\leq 3\sigma$ ($P_{\chi^2}$(obs)).  For all 30 Myr values of a given [$Q$,~$D$] pairing, we provide the means from the 20 individual simulations.  
For $M_c$ and $f_{\rm b}$(tot), we also provide the standard error on the mean, while for $\sigma_{\rm 0}$(obs) and $f_{\rm b}$(obs) we provide the upper and lower limits within which lie 95\% of our random 
sight lines across the 20 simulations.

\begin{figure}[!t]
\plotone{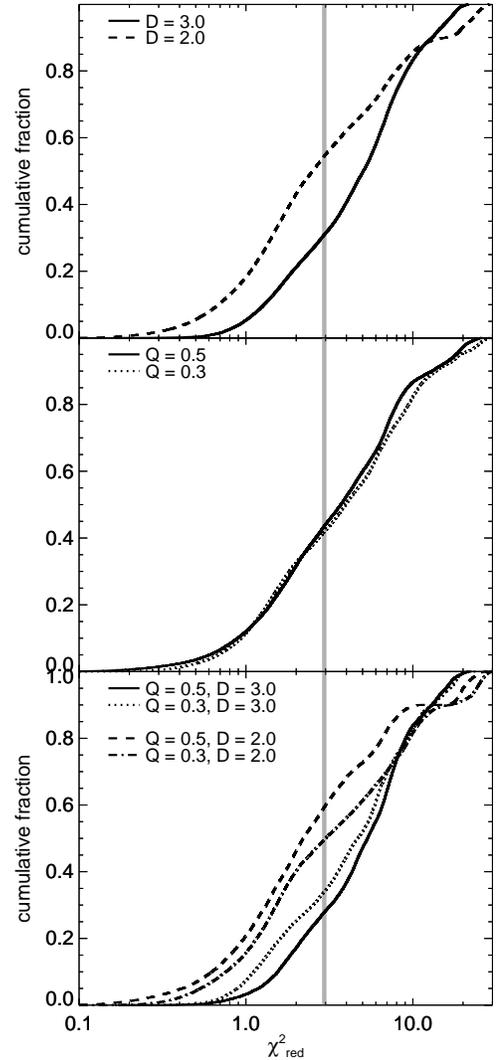}
\caption{
Distribution of $\chi^2$ values comparing the radial distribution of the binary frequency from the NGC~1805 $N$-body simulations to the \citet{li13} observations 
(shown in Figure~\ref{bfreqvr}).  We define the $\chi^2$ value as in Equation~\ref{chi2eq}, and perform this analysis for all 1000 sight lines to each of our simulations
with the given virial ratio ($Q$) and fractal dimension ($D$).  
The top and middle panels each combine the sets of simulations that either have the same fractal dimension (top) or virial ratio (middle), while the 
bottom panel shows each set of simulations separately.
The gray vertical line marks the $\chi^2$ value that indicates a $3\sigma$
difference between the observations and simulations (at the given number of degrees of freedom).  
\label{chi2fbvr}
}
\end{figure}

In Figure~\ref{bfreqvr}, we compare the observed radial distribution of the binary frequency from \citet{li13} to the same simulation shown in Figure~\ref{sdensity}.
Note the very close agreement in both the binary frequency and radial distribution.  Furthermore, this simulation reproduces the bimodal structure of the observed radial 
distribution of the binary frequency while drawing initial binaries from the \textit{same distributions} as our NGC~1818 model \citep{gel13b}, which has a distribution 
of binary frequency at the same chronological age that falls towards the core.  We discuss the reasons for this difference in Section~\ref{compto1818}.

Looking now at the full grid of NGC~1805 simulations, we compare the observed and simulated radial distributions of the binary frequency through a $\chi^2$ analysis, using 
the same procedure as in \citet{gel13b}.  For clarity, we repeat the definition here:
\begin{equation}
\chi^2 = \sum\limits_r \frac{(f_{\rm e}(<r)-f_{\rm o}(<r))^2}{e[f_{\rm o}(<r)]^2},
\label{chi2eq} 
\end{equation}
where $f_{\rm e}(<r)$ and $f_{\rm o}(<r)$ are the simulated and observed cumulative binary frequency inside the given radius, respectively, and $e[f_{\rm o}(<r)]$ is the error on the 
observed cumulative binary frequency at the same radius.  We have not constrained the simulation results, and therefore we have eight degrees of freedom (one for each bin in Figure~\ref{bfreqvr}).

The distributions of these $\chi^2$ values, combining results from each of the 1000 projections for each of the 20 models with
a given [$Q$,$D$] pairing, are shown in Figure~\ref{chi2fbvr}.  
Additionally, we provide the percent of sight lines through these simulations that match the observed radial distribution of the binary frequency to within 
$\leq3\sigma$ in Table~\ref{tab1}.
First, we find that the initially substructured simulations ($D = 2$) have a distribution of $\chi^2$ values shifted 
significantly to lower values than that of the initially smooth simulations ($D = 3$; top panel of Figure~\ref{chi2fbvr}).  A Kolmogorov-Smirnov test comparing these 
two distributions returns a probability $<10^{-7}$ that the two distributions are drawn from the same parent population.
Conversely, the distribution of $\chi^2$ values for the NGC~1805 simulations initially in equilibrium ($Q = 0.5$) cannot be distinguished from that of the initially collapsing simulations ($Q = 0.3$), as shown in the middle panel of Figure~\ref{chi2fbvr}.  
Thus we find a preference for substructured initial conditions in our models of NGC~1805 (which we also found for our NGC~1818 models in \citealt{gel13b}).

Interestingly though, only about half of the simulations reasonably reproduce the observed radial distribution of the binary frequency.
The gray vertical line shows the $\chi^2$ value for a 3$\sigma$ deviation of the simulation from the observations (at $\chi^2_{\rm red} \sim$ 2.95).
We can consider the $\sim$50\% of simulations with higher $\chi^2$ values than this limit to be poor fits to the observations.  
In general the models that fail to reproduce the observations do so either because they 
have a different total binary frequency or a radial distribution of the binary frequency that only decreases towards the core.  
This second point is in line with the results of \citet{gel13b}, where we find that, for smooth initial conditions, the simulations require at least one to two half-mass relaxation times 
to produce a bimodal distribution, and NGC~1805 has only lived through about one half-mass relaxation time, or less (see Section~\ref{compto1818}). 

This may indicate that there exists a more accurate choice of initial [$Q$,$D$], or other initial parameters, that best reproduces the observed cluster today.
However, our goal here is not to identify the exact initial parameters of the cluster, but to investigate whether both NGC~1805 and NGC~1818 could reasonably have been born 
with very similar primordial binary populations.
We find that we can closely reproduce the observations from a subset of our simulations, and therefore we conclude that the observations of both 
NGC~1805 and NGC~1818 can be reproduced from simulations beginning from very similar initial conditions (with appropriate modifications to the initial 
cluster mass and virial radius).
In the following section, we step back and examine the overall evolutionary trends of the binaries in our different NGC~1805 models.

\begin{figure}[!t]
\plotone{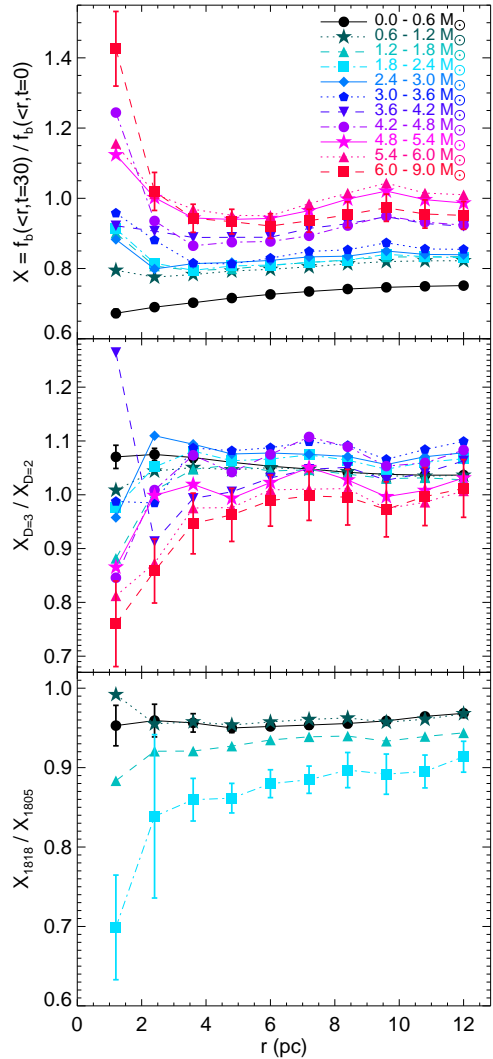}
\epsscale{1.}
\caption{
Distributions of binary frequency as a function of radial distance from the cluster center for binaries with different primary-star masses.
We cover the same radial domain as the observations, though with a slightly different bin size.  Each point 
represents an analysis of the stars inside the given radius.  However here we do not limit by mass ratio, and we take the three dimensional radius, rather than 
projected (as we did in Figure~\ref{bfreqvr} to compare with the observations).  
Different plotting styles and colors identify the different mass bins and are indicated in the top panel. 
Each point shows the mean value from each of the 20 simulations for the given [$Q$,$D$] pairing.  
Uncertainties show the standard errors on these means (with error propagation in the plotted ratios).
As an indicator of the range in uncertainties from all mass bins, and for ease of viewing, we only show uncertainties for the lowest and highest mass bin 
(which have approximately a factor of 200 difference in the total number of binaries).  
In the top panel, we show the binary frequency as a function of radius at 30 Myr (the age of NGC~1805) for our [$Q$,$D$] = [0.5,~2] NGC~1805 simulations, the set of 
simulations that are most consistent with the binary observations (e.g., see Figure~\ref{chi2fbvr} and Table~\ref{tab1}).
We normalize each bin by the value at the start of the simulation to remove any stochastic fluctuations resulting from drawing small numbers of stars from broad distributions 
in each of the individual simulations.  In the middle panel, we plot the ratio of these values to those of our [0.5,~3] NGC~1805 simulations, the set of simulations 
that are the least consistent with the binary observations.  Finally, in the bottom panel we plot the ratio 
of the values in the top panel to a similar analysis of our NGC~1818 simulations with [$Q$,$D$] = [0.5,~2] (the same initial [$Q$,$D$] as shown for NGC~1805 in the top panel).
\label{fbvrvmass}
}
\end{figure}

\section{Dependence of the Radial Distribution of the Binary Frequency on Stellar Mass} \label{phenom}

We can reproduce the two different trends in the observed radial distribution of the binary frequency in 
NGC~1805 and NGC~1818 through $N$-body simulations that begin with the same initial binary frequency and binary parameters drawn from the same initial distributions, 
at the same chronological age of 30 Myr.
Moreover, we do not need to invoke a different primordial binary population, or other changes to the primordial population between these two clusters (other than a difference in 
initial cluster mass and virial radius), to explain these observations.
We begin here to investigate the differences between these two clusters, starting with an analysis of the binary frequency as a function of stellar mass. 
In \citet{gel13b} we investigated the effects of binary disruption, two-body relaxation and mass segregation as a function of time.  
Examining the binary sample by stellar mass allows us to probe these effects at one age in the simulation, since higher mass stars are expected to undergo 
dynamical relaxation processes more quickly than lower mass stars. 

We focus on the NGC~1805 simulations with [$Q$,$D$]~=~[0.5,~2], as these models most closely reproduce the observed radial distribution of the binary frequency 
for the observed mass range (see Figures~\ref{bfreqvr}~and~\ref{chi2fbvr} and Table~\ref{tab1}).  We divide the model into mass bins of 0.6~\Msolar, approximately the mean stellar 
mass in the simulation at 30 Myr.  At masses greater than 10 times this mean mass, the sample size gets too small to continue this bin size (without very large 
uncertainties), and we therefore group these highest mass stars together into one bin.  (The maximum mass of a star in our simulations at 30 Myr is $\sim$9 \Msolar.)
Due to the relatively small number of stars in NGC~1805, it is unfortunately not possible to retrieve useful binary frequency radial distributions from the observations as a function of 
mass, due to the relatively large statistical uncertainties, and therefore here we only analyze the simulations.

We plot the radial distribution of the binary frequency in different mass bins in Figure~\ref{fbvrvmass}.  Here we do not project the model, and instead use the
3D radial position for each star.  We also take all binaries into consideration, rather than limiting to mass ratios $\geq0.55$ as we do for direct comparisons 
to the \citet{li13} observations.  Although we do not explicitly impose any differences in the binary population as a function of primary mass, we choose to normalize 
the results at 30 Myr to those at the start of our simulation to remove any stochastic fluctuations that result from drawing relatively small numbers of binaries from 
broad distributions (most relevant for the higher mass bins, which have smaller numbers of binaries).  

Focusing on the top panel of Figure~\ref{fbvrvmass}, we point out two general trends in the binary frequency that clearly vary with mass.  First, at an age of 30 Myr,
the overall binary frequency increases towards higher-mass stars.   Second, the binaries with higher-mass primaries show more dynamically evolved radial distributions.
Furthermore, a $\chi^2$ test comparing the radial distribution of the binary frequency for the lowest-mass binaries to that of any of the other mass bins returns a 
distinction at very high confidence ($>6\sigma$).

This first phenomenon can be understood by also looking at the distributions of semi-major axes for binaries of different masses.  In the top panel of Figure~\ref{avmass}
we show the mean semi-major axis within bins of the primary star's mass for both the [$Q$,$D$] = [0.5,~2] and [0.5,~3] models.  As discussed in detail in \citet{gel13b}
and also relevant here, the widest binaries are dynamically disrupted early in the simulations, which accounts for the overall drop in binary frequency seen in 
Figure~\ref{fbvrvmass}.  Figure~\ref{avmass} shows that the higher-mass binaries survive out to larger separations than the lower-mass binaries 
(although this is more apparent for the initially smooth model, and we will discuss this difference in Section~\ref{subdif}).  
As the semi-major axes for all binaries were drawn from the same (broad log-normal) distribution, this result indicates that a larger frequency of higher-mass binaries 
than lower-mass binaries are able to survive disruptions by an age of 30 Myr.  

The second trend that we point out above may be expected from two-body relaxation timescale arguments.
Returning to the top panel of Figure~\ref{fbvrvmass} we see that, for instance, the binaries with primary masses less than the mean mass have a binary 
frequency that decreases towards the core of the cluster.  Moving to bins containing higher and higher mass binaries, the distributions increase more and more 
towards the cluster center, and also begin to display a bimodal distribution.  We discussed this behavior in detail in \citet{gel13b} as a typical trend that 
a population of binaries may go through \textit{as a function of time}.  Here we note that even at one given time, the same trend can be observed by examining 
binaries of different masses.  Moreover, this shifting from a decreasing to increasing trend of binary frequency towards the cluster core depends on the relaxation timescale, 
which decreases towards higher-mass stars.

\subsection{Differences in the Evolution of the Smooth and Substructured Models} \label{subdif}

Our models with substructured initial conditions ($D = 2$) reproduce the observations more closely than those with smooth initial conditions (see Figure~\ref{chi2fbvr} and Table~\ref{tab1}).
Below, we point out a few key differences between these two sets of simulations which may help to explain the reasons behind this preference. 
For simplicity, we will focus on the simulations that begin initially in virial equilibrium ($Q=0.5$).

We start with the middle panel of Figure~\ref{fbvrvmass}, where we plot the ratio of radial distributions of the binary frequency from the [$Q$,$D$] = [0.5,~3] (initially smooth)
simulations over that of the [0.5,~2] (initially substructured) simulations.  For nearly all mass bins, this ratio decreases towards the cluster core.
Moreover, this ratio decreases more strongly towards the core as we consider binaries with higher mass primaries.  
A $\chi^2$ test distinguishes the distribution for lowest-mass binaries from that of the highest-mass binaries at $>4.5\sigma$ 
(and at $>3\sigma$ for the highest-mass bin compared to the 0.6~-~1.2~\Msolar\ bin, the 1.8~-~2.4~\Msolar\ bin and the 2.4~-~3.0~\Msolar\ bin).
In other words, at 30 Myr, the binary frequency in the core for a given binary primary mass is lower in the simulations with smooth initial conditions than for those with 
substructured initial conditions, and this difference is more pronounced for higher-mass binaries. 

\begin{figure}[!t]
\plotone{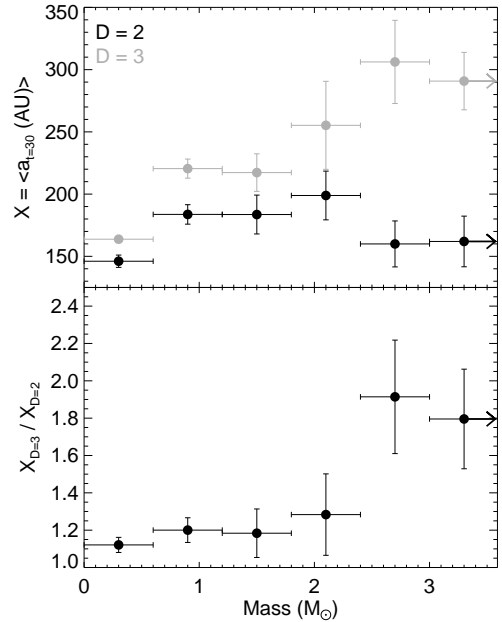}
\caption{
Distributions of mean binary semi-major axis for bins of different binary primary mass, comparing our NGC~1805 simulations with [$Q$,$D$] = [0.5,~2] and [0.5,~3] at 30 Myr.
We only include binaries with main-sequence primary stars within one initial half-mass radius ($\sim$3.07 pc) of the cluster center.
The points show the mean values from each of the 20 respective simulations for the given [$Q$,$D$] pairing, and the vertical error bars show the standard errors on 
the means (with error propagation in the bottom panel).
The horizontal error bars indicate the bin size; the last bin extends to the most massive stars in the cluster ($\sim$9~\Msolar), but is truncated here for ease of viewing.
In the top panel we compare the distributions for the two sets of simulations directly (with the $D$=2 simulations in black and the $D$=3 simulations in gray).
In the bottom panel we take the ratio of the results from these two simulations.
\label{avmass}
}
\end{figure}

It may be tempting to conclude that the lower binary frequency at 30 Myr in the core of the initially smooth models is due to the more efficient disruption of binaries in those models,
as compared to the initially substructured models.
However, turning to Figure~\ref{avmass}, where we compare the mean semi-major axes for binaries in bins of increasing primary mass at 30 Myr, we see that in fact the opposite is true.
Again, we began all of our simulations by drawing binaries from the same initial distribution of semi-major axes (with no dependence on initial masses) and 
allow dynamical encounters to naturally disrupt wide binaries.  The simulations that begin with initially substructured density distributions are more 
effective at disrupting binaries, of all primary masses.  
At an age of 30 Myr, the mean semi-major axis for binaries within one initial half-mass radius ($r_h(0) \sim 3.07$~pc) for the 20 simulations initially with
smooth density distributions ([$Q$,$D$]= [0.5,~3]) is $190~\pm~3$~AU, compared to the $160 \pm 3$~AU for the simulations with initially substructured density distributions 
([$Q$,$D$]= [0.5,~2]) within the same radial domain.
A Kolmogorov-Smirnov test comparing these two semi-major axis distributions returns a probability of 2$\times10^{-10}$ that the two are drawn from the same parent distribution.

This distinction is also seen in the top panel of Figure~\ref{avmass}, where the mean semi-major axis in every 
mass bin for the [0.5,~3] simulation is higher than that of the [0.5,~2] simulation, and in the bottom panel, where the ratio of these values is 
always greater than one.  The bottom panel of Figure~\ref{avmass} also shows that the ratio of the mean semi-major axes for these two sets of simulations 
is largest for binaries with the highest-mass primaries.
This result indicates that in the initially substructured simulations, the higher-mass binaries undergo more energetic encounters than in the initially smooth simulations.

These effects can all be understood under the same physical framework, as follows.  The initially substructured simulations undergo more rapid relaxation 
than the initially smooth simulations early in the cluster evolution.  
This situation physically resembles the conditions of violent relaxation \citep{lyn67,els87,spe15}, where the response of particles to processes like mass 
segregation is stronger if there are significant gradients in the background potential, as we also have here in the presence of substructure.
This causes the higher-mass binaries to segregate more quickly towards the core, producing the 
effect seen in Figure~\ref{fbvrvmass}.  
The result that star clusters with initially substructured density distributions undergo mass segregation more rapidly is well known from 
$N$-body star cluster models, and was first discussed by \citet{mcm07} (and soon thereafter by, e.g., \citealt{all09} and \citealt{moe09} using more extensive models),
as a potential explanation for
very young star clusters that are observed to have significant mass segregation, which is inconsistent with expectations from standard two-body relaxation processes 
(from unsegregated initial conditions).  We confirm this result here, now from the viewpoint of binary stars. 

Furthermore, because of pockets of initially very high density in the substructured simulations relative to the smooth simulations, the semi-major
axis distributions for binaries of all masses are shifted to lower values, by a factor of about 1.2, on average, inside of $r_h(0)$
(see also the top panel of Figure~\ref{avmass}).  
Additionally the higher-mass binaries in the substructured models segregate towards these high density regions
most efficiently, and therefore show an even larger difference in the semi-major axis distributions from the initially smooth simulations than do the 
lower-mass binaries (bottom panel of Figure~\ref{avmass}).  

\section{A Comparison to NGC~1818} \label{compto1818}

As presented in \citet{li13} and discussed in Section~\ref{intro} in this paper, unlike NGC~1805, the radial distribution of the binary frequency in the LMC star cluster NGC~1818
decreases (only) towards the core.  This is intriguing in its own right because older star clusters are consistently observed to have the opposite radial trend in binary frequency,
but is even more interesting when compared to NGC~1805, which is also located in the LMC and has the same chronological age, but instead shows a bimodal distribution in binary frequency 
with radius.  Timescale estimates from the observations indicate that the current half-mass relaxation time in NGC~1805 is 4 to 5 times shorter than in NGC 1818 \citep{deg02}.
We expand on this discussion below, by comparing our simulations and through simple analytic timescale estimates, to argue that NGC~1818 is less dynamically evolved than NGC~1805, and 
this can explain the difference in the radial distributions of their binary frequencies.  

We re-analyze the two [$Q$,$D$] = [0.5,~2] simulations of NGC~1818 from \citet{gel13b}, and compare the results to those of our 20 [0.5,~2] NGC~1805 simulations in the bottom
panel of Figure~\ref{fbvrvmass}.  Here we do not plot the distributions of the highest-mass binaries because the uncertainties become so large for the NGC~1818 simulations (due to small sample sizes) that
a precise comparison is not possible. First, we find that the overall binary frequency in NGC~1818 is lower.
This is due to the higher total cluster mass of NGC 1818, and therefore the higher velocity dispersion, which results in more binaries being disrupted \citep[see also][]{sol08}. 
More importantly for this discussion, the higher-mass binaries in NGC~1818 have a much lower binary frequency towards the core than in 
NGC~1805, while there is no such radial trend for the lower-mass binary comparison.  Formally, the distribution for highest-mass binaries can be distinguished from those of 
both the two lowest-mass binary bins at $>3\sigma$ confidence.  We interpret this in an analogous way to our interpretation of the middle panel of the same figure; NGC~1818 
is less dynamically evolved than NGC~1805 at the same chronological age. 

\begin{figure}[!t]
\plotone{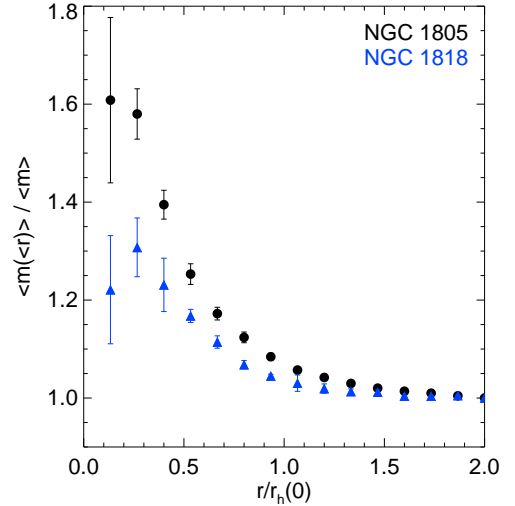}
\caption{
Mass distributions as functions of radius from the respective cluster centers in the NGC~1805 (black circles) and NGC~1818 (blue triangles) models.
Each point shows the mean mass inside of the given radius divided by the mean mass of all stars inside of two times the initial half-mass radius, $r_h(0)$, at 
a time of 30 Myr in the respective model.  For both NGC 1805 and NGC 1818, we use the [$Q$,$D$] = [0.5,~2] simulations (as we also show in other figures in this paper).
The uncertainties for the NGC~1805 points show the standard errors on the means across the 20 individual simulations.
For the uncertainties on the NGC~1818 points, we simply show the range in values from the 2 individual simulations.
In constructing this figure, we consider each binary as one object with a mass equal to the sum of the individual component masses.
\label{masseg}
}
\end{figure}

This difference in dynamical age is also apparent when we compare the radial mass distributions in the NGC~1805 and NGC~1818 models in Figure~\ref{masseg}.  
At the same chronological age of 30 Myr (and for the same [$Q$,$D$] = [0.5,~2] initial conditions), the radial mass distribution rises more steeply towards the core in 
the NGC~1805 model than in the NGC~1818 model.  
Moreover, the NGC 1805 model achieves a higher degree of mass segregation than the NGC~1818 model, within the same cluster lifetime.  

This result agrees with simple timescale arguments.  For instance, we calculate the initial half-mass relaxation time for stars with masses equal to the mean mass of 
an object (single or combined mass of a binary) in our NGC~1805 [0.5,~2] model to be about 50  Myr, while for the NGC~1818 [0.5,~2] models a similar calculation yields about 380 Myr
(following \citealt{spi87}, though the formula does not account for substructure).  
This result agrees roughly with \citet{deg02}, who estimate from observations that the present-day half-mass relaxation time in NGC 1805 is 4 to 5 times shorter than in NGC 1818.
Interestingly, a simple estimate of the crossing time for these two models yields about 5 Myr and 8 Myr, 
respectively.  Thus, the early phase of dynamical disruption, which creates a decreasing trend in binary frequency towards the core, likely operated on a similar timescale in both 
NGC~1805 and NGC~1818, but the recovery of the binary frequency in the core through mass segregation processes operates $\sim$~5 - 10 times faster in NGC~1805 than in NGC~1818.

\begin{figure}[!t]
\plotone{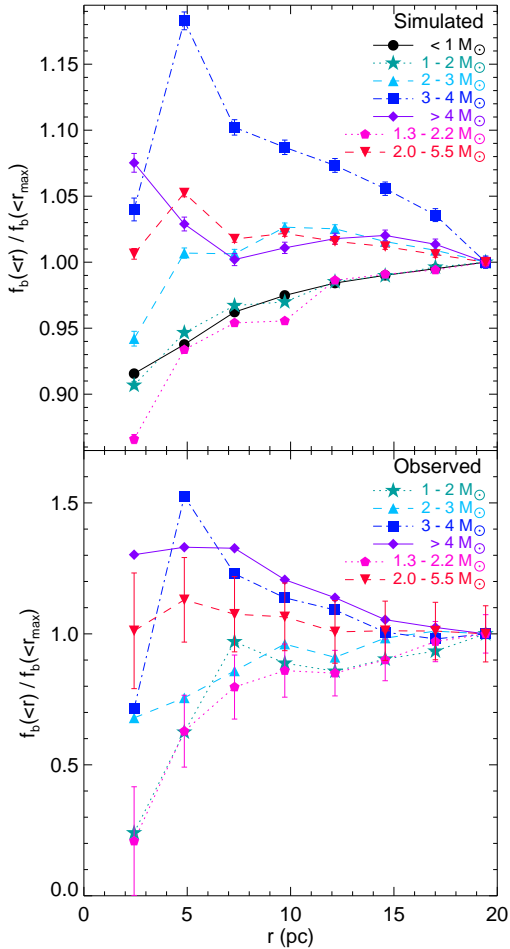}
\caption{
Comparison of the radial distributions of the binary frequency in different mass bins for one simulation of NGC~1818 with [$Q$,$D$] = [0.5,~2] from \citet[][top]{gel13b} and 
the observations from \citet[][bottom]{li13}.  The different plotting styles for the different mass bins are indicated on each plot.  
As in Figure~\ref{bfreqvr}, we show the binary frequency inside the given radius. We only count binaries with mass ratios $q = m_2/m_1 \geq 0.55$, and any binary with $q < 0.55$ 
is counted as a single star.  For the simulation, we use the same 1000 sight lines as in \citet{gel13b}; the points in the top panel show the 
means of these values, with uncertainties (visible when larger than the plotted symbols) indicating the standard error on these means. 
Uncertainties on the observations (derived using Poisson counting statistics, see \citealt{li13}) are much larger, and are shown in the bottom panel for two representative samples, 
for clarity.
\label{n1818comp}
}
\end{figure}

We also note that although NGC~1805 may be more dynamically evolved, NGC~1818 still exhibits a similar trend in the radial distribution of the binary frequency 
as we find in our NGC~1805 models when examining binaries of different mass bins.  In Figure~\ref{n1818comp} we show the result from re-analyzing one of the 
\citet{gel13b} NGC~1818 simulation with [$Q$,$D$] = [0.5,~2] (top), compared to the observations of NGC~1818 in the same mass bins.  
(Unlike NGC~1805, in NGC~1818 the larger number of stars 
enables us to divide the sample in different mass bins, though the uncertainties are large.)  Again, we see that the lowest mass binaries show a decreasing binary frequency 
towards the cluster core, while moving towards higher-mass binaries shows the trend shifting towards an increasing binary frequency towards the core 
(in both the observations and simulations).  

Finally, the result for NGC~1818 shown in Figure~\ref{n1818comp}, may help to explain the differences between the observations of \citet{li13} and \citet{els98}, who 
found opposite trends in the binary frequency as a function of radius.  \citet{li13} included binaries with primary masses between about 1.3~\Msolar\ and 2.2~\Msolar, and 
found a decreasing binary frequency towards the cluster center, while \citet{els98} included binaries with primary masses between 2~\Msolar\ and 5.5~\Msolar, and found an 
increasing binary frequency towards the cluster center.  We include these mass bins in Figure~\ref{n1818comp} (both for binaries with mass ratios $\geq 0.55$), 
and find different radial trends. We reproduce the observations of \citet[][as in \citealt{gel13b}]{li13} for the lower-mass sample, while the higher-mass binaries 
of the \citet{els98} sample show a flat, or perhaps marginally increasing, trend in binary frequency 
towards the core (similar to the conclusions from the $N$-body models of \citealt{els98}).  
\citet{deg13} and \citet{li13} speculated that this difference in binary mass may explain the discrepant observations, and we bolster this argument here.
We note that, as stated in \citet{gel13b}, our NGC~1818 models do not produce as high a binary frequency as observed by \citet[][using their observational
constraints in our analysis of the simulations]{els98}.  Therefore we do not perform a more detailed comparison to the observations.
We simply offer this discussion as a possible way to reconcile the analyses of \citet{els98} and \citet{li13}.

\section{Discussion} \label{discuss}

The two rich LMC star clusters NGC~1805 and NGC~1818 are both approximately the same chronological age, but are observed to have two different trends in the radial distribution
of their binary frequencies.  
We found in \citet{gel13b} that the radial distribution of the binary frequency can evolve over time from 
one that decreases towards the core (like in NGC~1818) to a bimodal distribution (like in NGC~1805) and eventually to a distribution that increases only towards the core 
(as observed in many older star clusters).  
The process occurs on a two-body relaxation timescale, and because NGC~1805 is less massive and more compact than NGC~1818, the expected relaxation time is shorter in NGC~1805.
Indeed \citet{deg02} estimated from observations that the present-day half-mass relaxation time in NGC~1805 may be 4 to 5 times shorter than in NGC 1818.
We therefore conjectured in \citet{gel13b} that if the binaries in both clusters follow the same evolutionary sequence, the different 
observed radial trends in the binary frequencies between these two clusters may be further evidence that NGC~1805 is more dynamically evolved than NGC~1818.  

We confirm this suggestion here through detailed $N$-body modeling of NGC~1805.  We can reproduce the observations of NGC~1805 through $N$-body models that draw from \textit{the same}
initial binary distributions, fractal dimension, and virial ratio as our NGC~1818 models (which reproduce the observations of that cluster), with the only differences 
being the initial masses of the clusters and the initial virial radii.  
Moreover, our simulations show that both clusters could have formed with binaries drawn from the same parent population, but are observed to be different today because of the 
clusters' different \textit{dynamical ages}
(where dynamical age is the number of relaxation times that the cluster has lived through).

Furthermore, we show above that binaries with different mass primaries undergo this evolutionary sequence of the radial distribution of the binary frequency at different rates, 
where the higher-mass binaries enter the bimodal and mass segregated phases earlier than the lower-mass binaries (e.g., top panel of Figure~\ref{fbvrvmass}).  This 
is expected from relaxation time arguments, and we show that this difference can also be used to compare the dynamical ages of clusters, for instance as we do in 
the bottom panel of Figure~\ref{fbvrvmass}.  This also highlights the importance of examining binaries in the same mass range when comparing observations across clusters.

As with our NGC 1818 models, we find that our NGC~1805 models that initially have some degree of substructure (here a fractal distribution with dimension $D=2$) 
more often match the observed radial trend in binary frequency.
We do not attempt to quantify the exact degree or nature of this substructure here.
Indeed it is doubtful that such an analysis would be possible, given the available observations, and that the substructure in our models is erased before an age of 30 Myr.
Nonetheless, the preference for substructure is encouraging, as this is consistent with observations of star forming regions \citep{lar95,car04,kra08,san09}.

Finally, we note again that, although we focus on an age of 30 Myr for both the NGC~1805 and NGC~1818 models, 
both \citet[][and references therein]{deg02} and \citet{li13} find the clusters to have marginally different ages, though these two papers each find a different cluster as being older.
\citet{li13}, who used improved isochrones and performed a more careful field star decontamination, find
a nominal age for NGC~1805 of 45 Myr and for NGC~1818 of 18 Myr.
This slight difference in age is in the right direction to help increase the expected difference in
the radial distributions of the binary frequency in both clusters (though if all else were equal between these clusters, this difference in age on its own would not be enough to 
produce the different radial trends in binary frequency).

\section{Conclusions} \label{conc}

The radial distribution of the binary frequency in a star cluster evolves with time due to dynamical disruptions from close encounters with other stars 
and mass-segregation processes, and can be used to track the dynamical age of a cluster.  
For a star cluster that is born with wide (soft) binaries, the early evolution of the binaries is dominated by disruptions,
which decrease the overall binary frequency, and establish a decreasing trend in binary frequency towards the cluster core, on approximately a crossing time.
The rich LMC star cluster NGC~1818 is observed in this phase of evolution.
On a two-body relaxation timescale, dynamical friction and mass segregation effects take over, causing the more massive binaries to sink towards the cluster core, 
which produces a bimodal radial distribution of the binary frequency.
NGC~1805 is observed in this phase of evolution.
Later on, as binaries towards the halo also begin to experience the effects of dynamical friction,
the binary frequency is transformed into one that only increases towards the core, as is observed in most older star clusters.
In \citet{gel13b} we showed that this evolutionary sequence can be tracked by looking at binaries of the same mass over time, and that the relevant timescale 
is not necessarily the chronological age, but instead the number of relaxation times the cluster has lived through, which we refer to as the cluster's dynamical age.

We show here that the same evolutionary sequence in the radial dependence of the binary frequency, can be observed at one chronological 
age for binaries of different primary masses.  The higher-mass binaries undergo dynamical friction and mass segregation processes at a faster rate than the lower-mass 
binaries, and therefore, although all binaries are subject to dynamical disruptions early on, the higher-mass binaries begin increasing their core binary frequency more quickly.

Our detailed $N$-body simulations confirm that NGC~1805 is dynamically older than NGC~1818.
Importantly, we can reproduce the observations of both clusters by drawing their stellar populations from the same parent population
(though starting from a different total cluster mass and concentration).
Furthermore, we show that today's observations of both LMC clusters can be reproduced by drawing their initial binary populations from 
distributions that are also consistent with observed solar-type binaries in the Milky Way field \citep{rag10} and observations of young Milky Way open clusters \citep[e.g. M35,][]{gel10}.  
These results are consistent with a hypothesis that the binaries in NGC~1805 and NGC~1818 were born with similar properties as those in Milky Way clusters, and 
suggest that binaries may form with similar distributions of orbital parameters and masses within a variety of different environments.
We suggest that the radial distributions of the binary frequencies in NGC~1805 and NGC~1818 are different today simply because we are catching them at slightly different stages along 
a very similar evolutionary sequence.

\acknowledgments
AMG is funded by a National Science Foundation Astronomy and Astrophysics Postdoctoral Fellowship under Award No.\ AST-1302765.
RdG and CL acknowledge partial research funding from the National Natural Science Foundation of China through grant 11373010.
AMG also fondly thanks Hannah Jeanne Trouille Geller for allowing him to run and monitor these simulations during her nap times.

\bibliographystyle{apj}
\bibliography{ngc1805.ms}

\end{document}